\documentstyle[12pt]{article}
\makeatletter
\if@twoside
\else \oddsidemargin 00pt \evensidemargin 00pt
\fi
\let\@eqnsel = \hfil

\expandafter\ifx\csname mathrm\endcsname\relax\def\mathrm#1{{\rm
#1}}\fi
\@ifundefined{mathrm}{\def\mathrm#1{{\rm #1}}}{\relax}

\makeatother

\begin{document}
\thispagestyle{empty}
\null
\hfill CERN-TH/95-188
\vskip 1cm
\vfil
\begin{center}
{\Large \bf Naturalness Constraints in Supersymmetric Theories
with Non-Universal Soft Terms
\par} \vskip 2.5em
{\large
{\sc S.\ Dimopoulos\footnote{\rm On leave of
absence from Stanford University, Stanford, Ca. 94305, USA.}} and {\sc
G.F.\ Giudice\footnote{\rm On leave of
absence from INFN, Sezione di Padova, Padua, Italy.}}
\\[1ex]
{\it Theory Division, CERN, Geneva, Switzerland} \\[2ex]
\par} \vskip 1em
\end{center} \par
\vskip 4cm
\vfil
\begin{center}
{\bf Abstract}
\end{center}
\par
In the absence of universality the naturalness upper limits on
supersymmetric
particle masses increase significantly. The superpartners of the two
light
generations can be much heavier than the weak scale without extreme
fine-tunings; they can weigh up to about 900 GeV --- or even up to 5
TeV,
if SU(5) universality is invoked. This supresses sparticle-mediated
rare processes and consequently ameliorates the problem of
supersymmetric flavor violations. On the other hand, even without
universality, the
 gluino and stop remain below about 400 GeV while the
charginos and neutralinos are likely to be accessible at LEP2.

\par
\vskip 2cm
\noindent CERN-TH/95-188 \par
\vskip .15mm
\noindent July 1995 \par
\null
\setcounter{page}{0}
\setcounter{footnote}{0}
\clearpage
The naturalness problem of the standard model with fundamental
Higgs particles \cite{hier} provides
{\it la raison d'\^etre} for supersymmetry at low energies
\cite{hiersusy}.
Indeed supersymmetry cures the problem if
the soft-supersymmetry breaking scale, which behaves as a cutoff for
quadratic divergences, lies below the TeV range \cite{hiersusy,dg}.
This
concept has been put on more quantitative grounds
by the studies presented in refs.
\cite{ellis,bg}, where upper bounds on physical supersymmetric
particle masses were obtained.
In this letter we wish to revisit
the analysis of ref. \cite{bg} and extend it to models with
non-universal boundary conditions at the unification scale.

We follow the
procedure proposed in ref. \cite{bg} and quantify the naturalness
criterion by requiring that
\begin{equation}
\left| \frac{a_i}{M_Z^2}
\frac{\partial M_Z^2}{\partial a_i}
\right| < \Delta ,
\label{nat}
\end{equation}
if no fine-tunings larger than $1/\Delta$ are allowed\footnote{Anderson
and Casta\~no \cite{anderson} have
recently proposed new naturalness
criteria which avoid sensitivity on steep functional dependences.
Since these new criteria lead to some dependence on the
allowed variable range, we prefer to use eq. (\ref{nat}) which is
an adequate fine-tuning measure in the case of cancellation among
different terms.}.
In eq. (\ref{nat})
the $a_i$ stand for all the soft supersymmetry-breaking parameters of
the
model which determine, via radiative symmetry breaking, the value of
$M_Z$.
We proceed by fixing one of the soft
supersymmetry-breaking parameters from the equations
\begin{equation}
\frac{M_Z^2}{2}=\frac{m_1^2-m_2^2\tan^2\beta}{\tan^2\beta -1},
\label{z}
\end{equation}
\begin{equation}
\sin 2\beta =\frac{2m_3^2}{m_1^2+m_2^2},
\label{t}
\end{equation}
where $m_1^2$, $m_2^2$, and $m_3^2$ are the usual mass parameters
in the Higgs potential, to be understood as functions of $a_i$.
Next we impose eq. (\ref{nat}) in the explicit form
\begin{eqnarray}
\left|\frac{2 a_i}{(\tan^2\beta -1)M_Z^2}\left\{
\frac{\partial{m_1^2}}{\partial a_i}-\tan^2\beta
\frac{\partial{m_2^2}}{\partial a_i}-\frac{\tan\beta}{\cos
2\beta}\times
\right. \right. \nonumber \\
\left. \left. \left( 1+\frac{M_Z^2}{m_1^2+m_2^2}\right) \left[
2\frac{\partial{m_3^2}}{\partial a_i}-\sin 2\beta \left(
\frac{\partial{m_1^2}}{\partial a_i}
+\frac{\partial{m_2^2}}{\partial a_i}\right)\right] \right\} \right|
< \Delta ,
\label{enat}
\end{eqnarray}
and scan over the values of $a_i$ which satisfy eq. (\ref{enat}).

Let us first consider the minimal supersymmetric standard model,
defined by
only five independent soft supersymmetry-breaking parameters at the
unification scale: a common scalar mass ($\tilde{m}$), a common
gaugino
mass ($M$), a Higgs superfield mass term ($\mu$), trilinear ($A$) and
bilinear ($B$) coupling constants. We perform a scanning over the
region of
these five parameters searching for points which satisfy the
naturalness
criterion, eq. (\ref{enat}), under the requirement of a
correct electroweak symmetry breaking, eq. (\ref{z}), the stability of
the
potential, and the experimental constraint that all new charged
particles
must be heavier than $M_Z/2$.

The procedure followed here slightly differs from the calculation done
in ref. \cite{bg}, as the electroweak symmetry breaking condition
is evaluated at the physical $M_Z$ value, instead of $M_Z=0$, and as
the actual upper bounds on the physical
supersymmetric particle masses are derived from eq. (\ref{nat}),
instead of
inferring them from the maximum values of the soft
supersymmetry-breaking
parameters at the unification scale.
Therefore
the bounds presented here are more stringent than those obtained in
ref.
\cite{bg} since, given a choice of supersymmetric parameters,
not all of them can simultaneously saturate their maximum values.
Also, our results are presented
as a function of the physical
top-quark pole
mass (as opposed to the top-quark Yukawa coupling $h_t$ \cite{bg})
defined
as:
\begin{equation}
m_t=h_t \frac{\sqrt{2}M_W}{g}\sin \beta (1+\frac{4}{3\pi}\alpha_s ) ,
\end{equation}
where the right-hand side is computed at the scale $Q^2=m_t^2$.
This
leads to the upper bounds on the supersymmetric particle masses shown
by the solid lines in fig. 1, for $\Delta=10$
({\it i.e.} no more than 10\% fine tuning), assuming the
unification mass $M_{GUT} =1.5\times 10^{16}$
GeV
and the unified gauge coupling constant $\alpha_{GUT}^{-1}=24.5$.
To obtain these bounds,
we have solved the relevant renormalization group equations
keeping
only one-loop contributions from gauge coupling constants and the
top-quark
Yukawa coupling. This approximation is inadequate when $\tan \beta$
approaches $m_t/m_b$, since the bottom-quark Yukawa coupling $h_b$ is
no longer negligible. We have checked that all bounds are achieved
for $\tan \beta << m_t/m_b$, and therefore we do not believe that the
inclusion of $h_b$-effects can sizeably modify our results.
We present the limits only for neutralinos, gluino, and lightest
chargino and stop for reasons that will be clear in the following.
We have included the one-loop QCD corrections on the pole
gluino mass in the $\overline{\rm MS}$ scheme \cite{glu}, but used
tree-level
expressions for chargino and neutralino masses.

The limit on the lightest stop rapidly increases below $m_t=170$ GeV,
since in this range of $m_t$,
the contribution proportional to $\tilde{m}$ in eq. (2) suffers
an approximate cancellation,
and the bound on $\tilde{m}$ disappears.
The limits on gluino, chargino, and neutralino masses shown
in fig. 1 are only weakly dependent on $m_t$, apart from the ``knee"
at $m_t<165$ GeV, a remnant of the above-mentioned approximate
cancellation.
Below this ``knee" we observe that the chargino mass is $m_{\chi^+}<80$
GeV and the lightest and next-to-lightest neutralino masses are
$m_{\chi^0}<40$ GeV, $m_{{\chi^0}'}<80$ GeV.
Therefore,
if minimal
low-energy supersymmetry describes the world with no more than 10\%
fine tuning, then LEP2 has great chances to discover it.

These limits are
very stringent largely as a consequence of the heavy top quark. The
large top-quark Yukawa coupling has the tendency to drive $m_2^2$
to large negative values. Only by choosing smaller values of the
soft-supersymmetry breaking parameters can one obtain the physical
value of $M_Z$, without increasingly precise fine tunings.

It is well known \cite{gamb} that the renormalization-group improved
tree-level potential has a strong scale dependence, which is partly
reabsorbed by the one-loop corrections to the effective potential.
Indeed it has been claimed \cite{cas} that these corrections
considerably relax the limits on supersymmetric particle masses
from the requirement of no fine-tuned top-quark Yukawa coupling,
as computed in ref. \cite{ross}. Although here we consider the
top-quark mass as an experimental datum and not as an input variable,
the question of scale dependence has to be addressed. We have therefore
included in our calculation all one-loop corrections to
the effective potential proportional to the top-quark Yukawa coupling,
with the complete mixings in the stop sector. The bounds obtained in
this way are about 20-30 \%  less stringent, as shown by the dashed
lines
in fig. 1.

It is interesting to understand whether the strong limits shown
by the solid lines in
fig. 1 depend on the specific assumptions of the form of the
soft supersymmetry-breaking terms at the unification scale. We
consider therefore the extreme case in which each possible
gauge-invariant soft
supersymmetry-breaking term at $M_{GUT}$ is an independent parameter.
We retain however the unification condition of a common gaugino
mass, largely justified by the success of gauge-coupling-constant
unification in supersymmetry \cite{dg}.

In practice, in the approximation of keeping only one-loop
renormalization
effects from gauge coupling constants and the top-quark Yukawa,
this generalization implies only
two new free parameters $a_i$ with respect to the previous case:
$m_1^2$ and $m_2^2$ evaluated at $M_{GUT}$\footnote{The boundary
condition
of the term coming from hypercharge D-term can be absorbed in the
definition of the boundary conditions for $m_1^2$ and $m_2^2$. The
effect of
this term will discussed below.}. We can then identify
$\tilde{m}^2$
with $(\tilde{m_{t_L}}^2+\tilde{m_{t_R}}^2)/2$,
the average of the left- and right-handed stop squared masses at
$M_{GUT}$,
which is the only other new combination of soft supersymmetry-breaking
terms appearing in the renormalization group equations.
Apart from having seven free parameters instead of five, the upper
bounds on supersymmetric particle masses are obtained with the
same procedure followed above. For simplicity we have not included here
the one-loop corrections to the effective potential.

Notice that in the non-universal case, within our approximation of
keeping only
leading one-loop effects, the parameter $m_1^2(M_{GUT})$ can be
infinitely large, without implying large fine tunings since, in this
limit, its value leaves $M_Z$ unaffected ($M_Z^2/2
\to m_3^2-m_2^2$ as $m_1^2 \to \infty$). This is
in contrast to the previous case
since, under universality conditions, $m_1^2 \to \infty$ implies
$m_2^2 \to \infty$\footnote{However, as mentioned above,
for a particular value of the top-quark Yukawa,
the dependence on $\tilde{m}$ disappears from the
expression of $M_Z$, and $\tilde{m} \to \infty$ does not imply a
large fine tuning.}. More generally, if the universality assumption
is dropped, all of the first two generations of squark
and slepton masses can
become infinitely large, within our approximation,
without causing fine-tuning problems in $M_Z$. Nevertheless, even if
some
of the soft supersymmetry-breaking parameters can become very large,
not all of the upper bounds on supersymmetric particles are lost.
For the particles considered in fig. 1,
the mass bounds in
the non-universal case, denoted by dot-dashed lines in the figure,
are of the same order of magnitude as in the
universal case. Actually they become much more constraining when
accidental cancellations occur among the soft-breaking parameters
correlated
by universal boundary conditions, as it is the case for the stop, when
$m_t
<170$ GeV, or for the gluino, when $m_t<160$ GeV, see fig. 1.
Light charginos and neutralinos
are a signature of fine-tuning-free low-energy supersymmetry,
independently of the specific assumptions on the soft
supersymmetry-breaking terms at the unification scale.

These considerations suggest to divide the supersymmetric particles
into
two species: the ``brothers of the Higgs", particles whose masses
strongly influence the running of the Higgs mass parameters and
consequently are severely constrained by the naturalness criterion; and
the ``cousins of the Higgs", particles which apparently are weakly
constrained by naturalness.

Neutralinos, charginos, gluinos, and stops are ``brothers of the Higgs"
and their mass bounds, shown in fig. 1, weakly depend upon the
model-dependent boundary conditions at $M_{GUT}$.
The scalar partner of the left-handed bottom quark is also a
``brother",
because of the weak SU(2) symmetry. On the other hand the partner of
the
right-handed bottom is not a ``brother", but it may become it in the
regime of large $\tan \beta$, where the bottom Yukawa coupling is no
longer negligible.
It may seem
that the gluino is a ``brother" only thanks to the unification
assumption
on gaugino masses. In reality, even if we relax this assumption, the
strong gluino mass upper bound persists. Indeed the dominant gaugino
contribution to the running of $m_2^2$ comes indirectly from the gluino
influence on the evolution of the stop mass parameters.

Let us now turn to discuss the mass bounds of the ``cousins of the
Higgs",
namely the first two generations of squarks and sleptons and, as
previously
discussed, some of the third generation sparticles, depending on the
$\tan \beta$ regime. In order to
obtain analytical expressions for their mass upper bounds, we simplify
our
procedure and use the following approximation. Notice that there exist
values of $\tan \beta$ such that the coefficients multiplying
$\frac{\partial m_1^2}{\partial a_i}$ and $\frac{\partial
m_3^2}{\partial a_i}$ in eq. (\ref{enat}) vanish. This suggests that
the best bounds will come from $\frac{\partial m_2^2}{\partial a_i}$,
whose coefficient in eq. (\ref{enat}) has a non-vanishing minimum.
We therefore convert eq. (\ref{enat}) into:
\begin{equation}
a_i^2<\frac{\Delta}{4}M_Z^2\left| \frac{1}{2a_i} \frac{\partial
m_2^2}{\partial a_i}\right|^{-1}.
\label{eenat}
\end{equation}
This expression is particularly convenient since, in the approximations
used below, the term inside the absolute value is independent of $a_i$.

The most important dependence in $M_Z$ on the ``cousins" comes from
a one-loop effect induced by the hypercharge D-term. Integrating the
renormalization group equations in the usual approximation, we find:
\begin{equation}
m_2^2=-\frac{1-Z_1}{22}~{\rm Tr}(\bar{m}^2_Q+\bar{m}^2_{\bar{D}}-2
\bar{m}^2_{\bar U}-\bar{m}^2_L+\bar{m}^2_{\bar E}) +{\hat m}_2^2 ,
\label{dt}
\end{equation}
\begin{equation}
Z_1=\left(1+\frac{33}{20
\pi}\alpha_{GUT}\log\frac{M_{GUT}^2}{Q^2}\right)^{-1} =0.4~,
\end{equation}
where $\bar{m}^2_A$ ($A=Q,\bar{U},\bar{D},L,\bar{E}$) are the boundary
conditions of the soft-breaking masses of squarks and sleptons at
$M_{GUT}$,
all of which are independent parameters, and the trace in eq.
(\ref{dt})
is taken over generation space. The term in eq. (\ref{dt}) denoted by
${\hat m}_2^2$ contains the usual dependence on the other soft
supersymmetry-breaking terms.

Using eqs. (\ref{eenat}) and (\ref{dt}), we obtain the
bound\footnote{Of
course heavy ``cousins" require a certain degree of fine tuning not
only to
keep the Higgs light, but to keep the ``brothers" light as well.}:
\begin{equation}
\bar{m}_A<\sqrt{\frac{11\Delta}{2(1-Z_1)}}M_Z=900\sqrt{\frac{\Delta}{10}
}
{}~{\rm GeV}. \label{lim}
\end{equation}
This is approximately also the bound on the physical squark and slepton
masses, since the renormalization effects are at most 10{\%}, given the
naturalness bounds on gaugino masses.

In theories with universal boundary conditions, the hypercharge D-term
vanishes, as a result of anomaly cancellation. Actually it vanishes
also
in theories where the soft terms satisfy simple GUT relations like:
\begin{equation}
\bar{m}^2_Q=
\bar{m}^2_{\bar U}=
\bar{m}^2_{\bar E}\equiv
\bar{m}^2_{10}~,~~~~~~~~~~~~
\bar{m}^2_{\bar D}=
\bar{m}^2_L\equiv
\bar{m}^2_{\bar 5}~.
\label{rel}
\end{equation}
We refer to eqs. (\ref{rel}) as ``SU(5) universality".
Although it has been shown \cite{dp} that integration of heavy GUT
particles
can modify these relations already at the tree-level, eqs. (\ref{rel})
are still plausible conditions in certain GUT models. If they hold,
in order to bound the masses of the ``cousins of the Higgs", we have
to rely on two-loop effects or one-loop effects suppressed by small
Yukawa couplings.

We include these small effects as perturbations over the leading-order
solutions. Let us take for simplicity the case in which all trilinear
soft terms are zero, and write the solutions of the renormalization
group equations for the stop and Higgs mass parameters as:
\begin{equation}
m_i^2=\hat{m}_i^2+\delta m_i^2,
\end{equation}
where $\hat{m}_i^2$ are the solutions of the one-loop equations
including only gauge and top Yukawa couplings. Similarly, the
corresponding $\beta$-functions are separated into the part which
depends on gauge and top Yukawa couplings ($\hat{\beta}_i$) and
the perturbation ($\delta \beta_i$). At leading order, the evolution
equation for $\delta m_i^2$ is
\begin{equation}
\frac{d}{dt} \delta m_i^2 =\sum_j \frac{\partial \hat{\beta}_i}
{\partial m_j^2} \delta m_j^2 + \delta \beta_i \label{run}
\end{equation}
\begin{equation}
t\equiv \log \frac{M_{GUT}^2}{Q^2}.
\end{equation}

Let us first choose as perturbation the two-loop dependence on
$\bar{m}^2_{10}$ and $\bar{m}^2_{\bar{5}}$ not suppressed by small
Yukawa couplings. Integrating eq. (\ref{run}) using the appropriate
two-loop $\beta$-functions \cite{two}, we find:
\begin{eqnarray}
\delta m_2^2=\frac{\alpha_{GUT}}{4 \pi} \left[ \left(
\frac{1}{6}\log\frac{Z_3}{Z_1}-
\frac{9}{56}\log\frac{Z_2}{Z_1}-\frac{1-Z_1}{66}\right)
{\rm Tr}({\bar m}^2_{10}-{\bar m}^2_{\bar 5})\right. \nonumber \\
\left. +\left(
\frac{3}{2}Z_2+\frac{1}{22}Z_1-\frac{86}{99}-\frac{67}{99}
\xi +\frac{1-\xi}{2} G\right) {\rm Tr}(3{\bar m}^2_{10}+{\bar
m}^2_{\bar 5})
\right] \label{tl}
\end{eqnarray}
\begin{equation}
G=\frac{\int_0^t dt Z_3^{\frac{16}{9}}Z_2^{-3}Z_1^{-\frac{13}{99}}
\left( \frac{16}{9}Z_3 -3Z_2-\frac{13}{99}Z_1 \right)}
{\int_0^t dt Z_3^{\frac{16}{9}}Z_2^{-3}Z_1^{-\frac{13}{99}}}
\end{equation}
\begin{equation}
Z_i =\frac{\alpha_i (t)}{\alpha_i (0)},~~~~~~~\xi = \left(
1-\frac{m_t^2}{M_{FP}^2\sin^2\beta}\right),
\end{equation}
where $M_{FP}=192$ GeV is the top-quark infrared fixed point value.
Using eqs. (\ref{eenat}) and (\ref{tl}), we obtain the upper bounds,
which for $m_t$ varying in the range between 160 and 190 GeV vary in
the range
\begin{equation}
{\bar m}_{10}< 1.7 - 2.2 ~\sqrt{\frac{\Delta}{10}}~{\rm TeV}~,~~~~~~
{\bar m}_{\bar 5}< 3.6 - 5.6 ~\sqrt{\frac{\Delta}{10}}~{\rm TeV}~,
\end{equation}
where the more stringent limit corresponds to the heavier top quark
mass.

It is straightforward to repeat the calculation including as
perturbation
the one-loop dependence on Yukawa couplings different from the top one.
In this case one finds bounds which are weaker that those
obtained from the two-loop gauge effects.

In conclusion, the naturalness criterion sets very stringent bounds
on the masses of the ``brothers of the Higgs", particles whose
existence strongly influence the evolution of the Higgs mass
parameters. These bounds do not significantly depend on the particular
boundary conditions of the supersymmetry-breaking terms at $M_{GUT}$.
Neutralinos, charginos, gluinos, and stops belong to this class of
particles. The new experimental evidence of a heavy top quark has
made these bounds particularly constraining, suggesting that these
particles should be soon discovered, independently of the particular
model-dependent assumptions.

 On the other hand, bounds on the masses of the ``cousins of the
Higgs" -- first two generations of sleptons and squarks -- depend much
more on model assumptions. If one relaxes the usual universality
assumption,
their naturalness mass bounds can be significantly altered.
Nevertheless,
we have shown that a 10 \% fine tuning criterion still requires them
to be lighter than about 900 GeV, if there is no cancellation of
the hypercharge D-term contribution, eq. (\ref{dt}). If this
cancellation
occurs, maybe caused by GUT universality relations, two-loop
effects still provide
limits in the range between 2 and 5 TeV.
Although this ameliorates the problem of supersymmetric flavor
violations
\cite{mah}, it does not completely solve it. A certain degree of
degeneracy between sparticles or alignment of sparticle and particle
masses \cite{nir} is still needed to adequately suppress all flavor
violations.

\bigskip

\bigskip

\noindent{\large{\bf Figure Caption}}

\bigskip
\noindent {\bf Fig. 1.} Upper bounds on gluino, lightest and
next-to-lightest neutralino, and lightest chargino and stop masses
based
on the requirement of no fine tuning larger than 10 {\%}, {\it
i.e.} $\Delta =10$ in eq. ({\ref{nat}). The mass bounds approximately
scale as $\sqrt{\Delta}$. The solid (dashed) lines refer
to the minimal supersymmetric standard model with universal
boundary conditions at $M_{GUT}$ for the soft supersymmetry-breaking
terms, without (with) the inclusion of the one-loop effective
potential.
The dot-dashed lines show the mass upper limits for
non-universal boundary conditions at $M_{GUT}$, without the inclusion
of the
one-loop effective potential.
\end{document}